\documentclass[letter]{aa} 

\usepackage{natbib}
\bibpunct{(}{)}{;}{a}{}{,}

\usepackage{graphicx}
\usepackage{revsymb}	
\usepackage{amsmath}
\usepackage{txfonts}
\usepackage{amssymb}
\usepackage{geometry} 
\usepackage[parfill]{parskip} 
\usepackage{slantsc}
\usepackage{array}
\usepackage{amsfonts}
	
\usepackage{epstopdf}	
\usepackage{epsfig}	

\usepackage{url}

\usepackage[colorlinks=true,linkcolor=black, urlcolor=black, citecolor=black]{hyperref}

\usepackage{color}
\usepackage{xcolor}
\colorlet{rouge}{red!70!darkgray}

\usepackage{xfrac}

\usepackage{sidecap}
\usepackage[font=small, labelfont=bf]{caption}
\usepackage{subcaption}

\usepackage{threeparttable}

\begin{document}
      \title{Testing angular momentum transport processes with asteroseismology of solar-type main-sequence stars}

\author{J. B\'{e}trisey\inst{1} \and P. Eggenberger\inst{1} \and G. Buldgen\inst{1} \and  O. Benomar\inst{2,3} \and M. Bazot\inst{4} }
\institute{Observatoire de Genève, Université de Genève, Chemin Pegasi 51, 1290 Versoix, Suisse\\email: 	\texttt{Jerome.Betrisey@unige.ch}
\and National Astronomical Observatory of Japan, Solar Science Observatory, 2-21-1 Osawa, Mitaka, Tokyo 181-8588, Japan
\and  New York University Abu Dhabi, Center for Space Science, PO Box 129188, UAE
\and Heidelberg Institute for Theoretical Studies (HITS gGmbH), Schloss-Wolfsbrunnenweg 35, 69118 Heidelberg, Germany}
\date{\today}

\date{\today}

\abstract{Thanks to the so-called photometry revolution with the space-based missions CoRoT, \emph{Kepler}, and TESS, asteroseismology has become a powerful tool to study the internal rotation of stars. The rotation rate depends on the efficiency of the angular momentum (AM) transport inside the star, and its study allows to constrain the internal AM transport processes, as well as improve our understanding of their physical nature.}
{We compared the ratio of the rotation rate predicted by asteroseismology and starspots measurements of solar type stars, considering different AM transport prescriptions, and investigated if some of these prescriptions could observationally be ruled out.}
{We conducted a two steps modelling procedure of four main-sequence stars from the \emph{Kepler LEGACY} sample, which consists in an asteroseismic characterisation that serves as a guide for a modelling with rotating models including a detailed and coherent treatment of the AM transport. The rotation profiles derived with this procedure are used to estimate the ratio of the mean asteroseismic rotation rate with the surface rotation rate from starspots measurements for each AM transport prescriptions. Comparisons between the models are then conducted.}
{In the hotter part of the Hertzsprung-Russell (HR) diagram (masses typically above $\sim 1.2M_\odot$ at solar metallicity), models with only hydrodynamic transport processes and models with additional transport by magnetic instabilities are found to be consistent with measurements reported by \citet{Benomar2015} and \citet{Nielsen2017} who observed a low degree (below 30\%) of radial differential rotation between the radiative and convective zones. For these stars, which constitute a significant fraction of the \emph{Kepler LEGACY} sample, combining asteroseismic constraints from splittings of pressure modes and surface rotation rates does not allow to conclude on the need for an efficient AM transport in addition to the sole transport by meridional circulation and shear instability. Even the model assuming local AM conservation cannot be ruled out. In the colder part of the HR diagram, the situation is different due to the efficient braking of the stellar surface by magnetised winds. We find a clear disagreement between the rotational properties of models including only hydrodynamic processes and asteroseismic constraints, while models with magnetic fields correctly reproduce the observations, similarly to the solar case.}
{There is a mass regime corresponding to main-sequence F-type stars for which it is difficult to constrain the AM transport processes, unlike for hotter, Gamma Dor stars or colder, less massive solar analogs. The comparison between asteroseismic measurements and surface rotation rates enables to easily rule out the models with an inefficient transport of AM in the colder part of the HR diagram.}

\keywords{Stars: rotation -- Stars: interiors -- asteroseismology -- Stars: magnetic field -- Stars: fundamental parameters -- Stars: individual: KIC8006161, KIC8379927, KIC9139151, KIC12258514}

\maketitle

\section{Introduction}
Oscillations at the surface of stars carry an information about the stellar structure and their study permits to constrain the transport processes occurring inside the star, as well as characterise its rotation. These studies were first dedicated to helioseismology, because of the required data quality, and tremendous successes were achieved. For example, it was shown that the radiative interior of the Sun rotates nearly uniformly \citep[see e.g.][]{Schou1998,Thompson2003,Eff-Darwich&Korzennik2013}. Solar models computed with only hydrodynamic transport processes in radiative zones, such as meridional circulation and shear instability, were then found to predict a high contrast between core and surface rotation rates, in disagreement with helioseismic measurements \citep{Pinsonneault1989,Chaboyer1995,Eggenberger2005mag,Charbonnel&Talon2005}. Another efficient angular momentum (AM) transport process must then be operating in the solar radiative zone. Different candidates have been invoked for such an efficient AM transport in the Sun: internal gravity waves \citep[e.g.][]{Zahn1997,Charbonnel&Talon2005}, large-scale fossil magnetic fields \citep[e.g.][]{Mestel&Weiss1987,Charbonneau1993,Ruediger&Kitchatinov1996,Gough&McIntyre1998} and magnetic instabilities \citep[e.g.][]{Spruit2002,Eggenberger2005mag,Eggenberger2019}. The latter recently demonstrated that it could provide an interesting explanation to the helioseismic measurements of the internal rotation of the Sun simultaneously to the surface abundances of lithium and helium \citep[][]{Eggenberger2022}.

The recent development of space-based photometry missions, such as CoRoT \citep{Baglin2009}, \emph{Kepler} \citep{Borucki2010}, and TESS \citep{Ricker2015} in the last two decades enables to apply these studies to asteroseismology as well. During almost all the phases in the life of a star, the core contracts and the envelope expands, creating differential rotation with a core rotating faster than the envelope. Moreover, braking of the surface by magnetized winds can create radial differential rotation in solar-type stars with a convective envelope deep enough to host a dynamo. This trend can however be mitigated by an efficient AM transport. Key observational constraints have been obtained for subgiant and red giant stars with the asteroseismic determination of the core rotation rates for a large sample of these evolved stars \citep{Beck2012,Deheuvels2012,Deheuvels2014,Deheuvels2015,Deheuvels2017,DiMauro2016,Dimauro2018,Mosser2012,Gehan2018,Fellay2021}. Comparisons with rotating models have then revealed the need for an efficient AM transport mechanism in addition to meridional circulation and transport by the shear instability \citep{Eggenberger2012rg,Ceillier2013,Marques2013,Eggenberger2017,Eggenberger2019,Moyano2022}. Detailed asteroseismic studies of the internal rotation for some main-sequence (MS) stars, in particular for $\gamma$ Dor pulsators, also suggested that an efficient transport of AM is operating in the radiative zones of these stars similarly to the conclusion obtained for the Sun and evolved stars \citep[e.g.][]{Kurtz2014,Saio2015,Murphy2016,Ouazzani2019,Li2020,Saio2021}. An important question is related to the internal transport of AM for MS stars less massive than the $\gamma$ Dor pulsators, so typically for stars with masses lower than about 1.5 $M_\odot$. In this context, \citet{Benomar2015} (hereafter OB15) studied 22 MS solar-type stars observed by \emph{Kepler} and found that the average rotation rates deduced from asteroseismic measurements for these stars are very similar to their surface rotation rates. \citet{Nielsen2017} (hereafter MN17) reached the same conclusion using an independent approach based on two zones models fittings of the power spectrum. For the five \emph{Kepler} targets considered in their work, they found that the radial differential rotation did not exceed 30\% between the radiative and convective zones. 

In this study, we investigate how these observations can constrain the internal transport of AM in solar-type stars and shed some light on the physical nature of this transport. We considered two AM transport prescriptions, either with or without including magnetic Tayler instability. We examined how these prescriptions affect internal and surface rotation rates. We give a semi-quantitative assessment of their compatibility with existing measurements from OB15 and MN17, and whether some scenarios for AM transport can be ruled out. Although our study is based on synthetic models, we still used an advanced modelling to generate `realistic' models for the comparisons, whose structure reproduces the classical and (non-rotating) seismic constraints of an actual observed target. We  selected four solar-type MS stars from the \emph{Kepler LEGACY} sample \citep{Lund2017}, that we divided in two categories: Arthur, Barney, and Carlsberg, representative of the hotter region of the Hertzsprung-Russell (HR) diagram, and Doris, representative of the colder regions of the HR diagram (see Fig. \ref{fig:hr}). This selection is based on the size of the convective envelope that impacts the efficiency of surface braking. For our hottest targets, the convective envelope is shallow and inefficient braking is expected for masses above $\sim 1.2M_\odot$ at solar metallicity \citep{Kraft1967}. For Arthur, Barney, and Carlsberg, which lie close to this threshold, it is less clear how magnetic braking is behaving. We assumed a likely inefficient braking for these targets, and then discussed the relevance of this hypothesis. In Sect. \ref{sec:modelling_and_rotation_profiles}, we describe the asteroseismic modelling procedure and the physical input of the models. In Sect. \ref{sec:surface_rotation}, we compare the rotational properties of these different models to the available observational constraints, while the conclusions are given in Sect. \ref{sec:conclusions}.


\section{Stellar models}
\label{sec:modelling_and_rotation_profiles}

We summarised the observational data in Table \ref{tab:obs_constraints}. The luminosity was estimated from the spectroscopic parameters using the same procedure as for Kepler-93 in \citet{Betrisey2022} (hereafter JB22), but with distances from \citet{Bailer-Jones2021} and based on the parallaxes\footnote{The GAIA data of Arthur is flagged as unreliable. The estimated observed absolute luminosity should be considered with caution and was not reproduced by our models.} measured by \citet{Gaia2021}. The frequencies come from \citet{Lund2017} for Barney, Carlsberg, and Doris, and \citet{Roxburgh2017} for Arthur. 
\begin{table}[t!]
\centering
\caption{Observed and modelled data of Arthur, Barney, Carlsberg, and Doris.}
\resizebox{\linewidth}{!}{
\begin{tabular}{lcccccc}
\hline 
  & Unit & Arthur & Barney & Carlsberg & Doris & Ref.\\ 
\hline \hline
\textit{Observed data} \\
KIC & & 8379927 & 12258514 & 9139151 & 8006161 & \\ 
HD & & 187160 & 183298 & - & 173701 & \\ 
$T_{\mathrm{eff}}$  & (K) & $6067\pm 150$ & $5964\pm 60$ & $6043\pm 100$ & $5488\pm 100$ & 1\\ 
$\rm [Fe/H]$  & (dex) & $-0.10\pm 0.15$ & $0.00\pm 0.10$ & $0.05\pm 0.10$ & $0.34\pm 0.10$ & 1\\ 
$L$  & $(L_\odot)$ & $2.24\pm 0.12$ & $2.95\pm 0.11$ & $1.60\pm 0.06$ & $0.69\pm 0.03$ & 2\\ 
$\nu_{max}$  & ($\mu$Hz) & $2795.3\pm 6.0$ & $1512.7\pm 3.1$ & $2690.4\pm 11.8$ & $3574.7\pm 11.0$ & 3\\ 
$\Omega_{\mathrm{surf}}$ & (days) & $17.59\pm 0.36$ & $15.00\pm 1.84$ & $10.96\pm 2.22$ & $29.79\pm 3.09$ & 4\\
\hline 
\textit{Modelled data} \\
$M$ & ($M_\odot$) & 1.231 & 1.270 & 1.186 & 1.007 & \\
$R$ & ($R_\odot$) & 1.159 & 1.611 & 1.162 & 0.937 & \\
Age & (Gyr) & 1.46 & 4.04 & 2.03 & 5.55 & \\
$X_0$ & & 0.740 & 0.674 & 0.734 & 0.711 & \\
$Z_0$ & & 0.0201 & 0.0273 & 0.0169 & 0.0303 & \\
\hline
\end{tabular} 
}
\label{tab:obs_constraints}
{\par\small\justify\textbf{References.} (1) \citet{Lund2017} for Arthur, Barney, and Doris, and \citet{Furlan2018} for Carlsberg; (2) this work; (3) \citet{Lund2017}; (4) from starspots measurements, mean and standard deviation of \citet{Nielsen2013}, \citet{Reinhold2013}, and \citet{McQuillan2014} for Arthur, OB15 for Barney and Carlsberg, and \citet{Garcia2014} for Doris \par}
\end{table}

\begin{figure}[t!]
\centering
\includegraphics[scale=0.58]{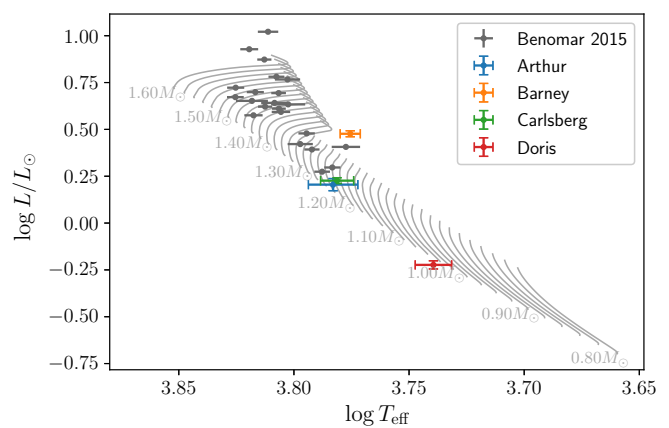} 
\caption{HR diagram of Arthur, Barney, Carlsberg, Doris, and of the sample of OB15. The tracks correspond to a grid slice with an initial chemical composition of $X_0=0.74$ and $Z_0=0.018$, and no overshooting.}
\label{fig:hr}
\end{figure}

The modelling procedure is divided in two main steps, described in details in Appendix \ref{app:detailed_modelling_procedure}. The first step, which consists in fitting the seismic information with a Markov Chain Monte Carlo (MCMC) in a grid of non-rotating models, serves as a guide for the second step, where we derive the rotation profiles using rotating models with a detailed and coherent treatment of the AM transport. The modelling procedure of the first step is similar to that of JB22; we used a grid of non-rotating models computed with the Code Li\'{e}geois d’\'{E}volution Stellaire \citep[CLES,][]{Scuflaire2008b}, and the frequencies were computed with the Li\`{e}ge Oscillation Code \citep[LOSC,][]{Scuflaire2008a}. The physical ingredients are the same as in JB22 (see Sec. 2.1). The minimisations were conducted with AIMS \citep{Rendle2019}, with a procedure that combines a mean density inversion \citep{Reese2012} with a fit of frequency separation ratios. This modelling approach provides robust stellar seismic models \citep[see e.g.][]{Buldgen2019f,Betrisey2022}, whose properties are given in Table~\ref{tab:obs_constraints}.

\begin{figure*}[t!]
\centering
\includegraphics[scale=0.445]{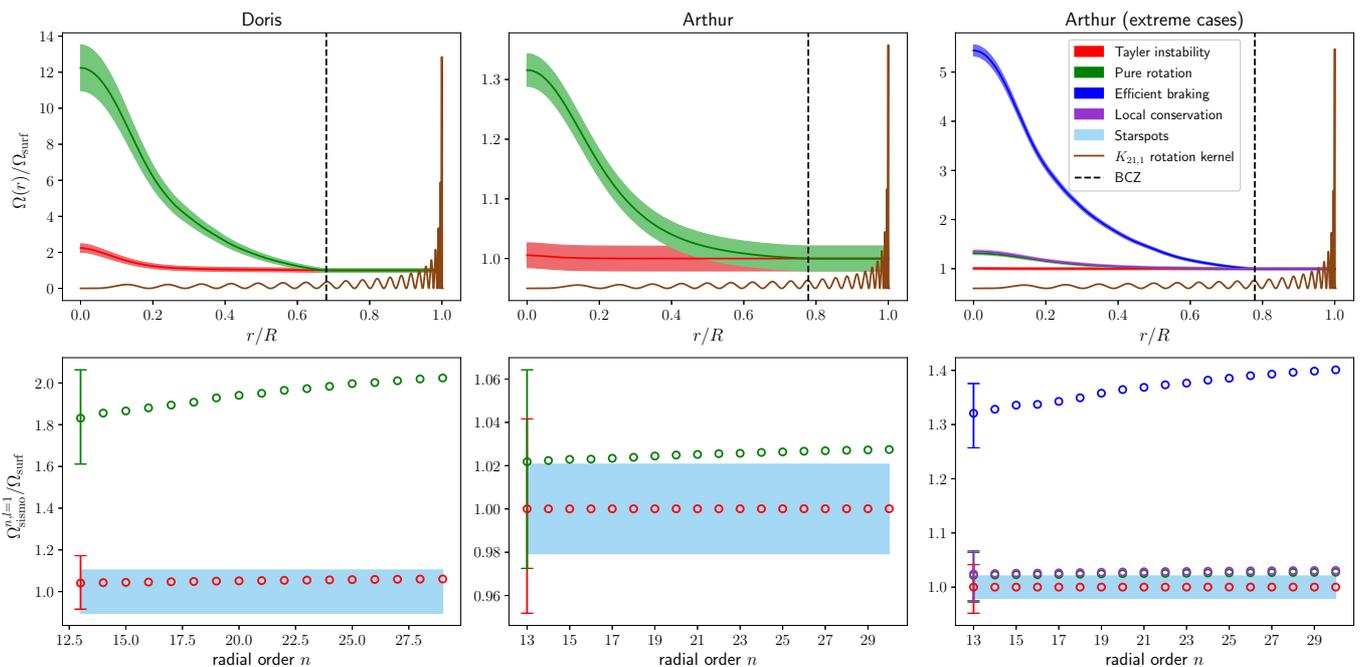} 
\caption{\textit{Top line:} Rotation profiles of Doris (left panel) and Arthur (center and right panels). The base of the convective zone (BCZ) is shown in dashed black. The rotation kernel around the $\nu_{max}$, namely $K_{21,1}$, is shown in brown, and is rescaled and shifted vertically for illustration purposes. \textit{Bottom line:} Ratio of the rotation rate predicted by asteroseismology and starsports measurements for Doris (left panel) and Arthur (center and right panels). The open circles are the theoretical predications, considering different AM transport prescriptions, accounting only for hydrodynamic processes (green circles), or for hydrodynamic processes and magnetic Tayler instability (red circles). The shallow convective envelope of Arthur likely implies an inefficient braking. For Arthur, we also tested extreme cases (right panels), by considering an AM transport with an efficient braking (blue circles) or assuming local AM conservation (purple circles). The error bar corresponds to the precision of the average observational rotational splitting measured by OB18, and the blue area to a solid-body rotation profile with the observational uncertainty of the surface rate from starspots measurements.}
\label{fig:figure2}
\end{figure*}

Once the targets properties are determined thanks to this detailed asteroseismic modelling, rotating models are computed with the Geneva stellar evolution code \citep[GENEC,][]{Eggenberger2008}. The computation of these rotating models is first based on the initial parameters obtained from the asteroseismic modelling and these parameters are then adjusted to correctly reproduce the stellar properties given in Table 2. The GENEC code assumes shellular rotation \citep{Zahn1992}, and the internal AM transport is computed along the stellar evolution, by accounting for shear instability, meridional circulation, and AM transport by magnetic instability as in \citet{Spruit2002}. The advecto-diffusive AM transport in the radiative zone is described by:

\begin{small}
\begin{align}
\rho \frac{d}{dt}\left(r^2\Omega\right)_{M_r} = \frac{1}{5r^2}\frac{\partial}{\partial r}\left(\rho r^4 \Omega U(r)\right) + \frac{1}{r^2}\frac{\partial}{\partial r} \left[(D_{shear}+\nu_{\rm TS})\rho r^4\frac{\partial\Omega}{\partial r}\right],
\end{align}
\end{small}\unskip
where $\rho$ is the mean density, $r$ is the radius, $\Omega$ is the mean angular velocity on an isobar, and $U$ is the radial component of the meridional circulation. The AM transport by shear instability is described by the coefficient $D_{shear}$ following \citet{tal97}, and the $\nu_{\rm TS}$ is the diffusion coefficient corresponding to the transport by the Tayler-Spruit dynamo \citep[see e.g.][]{Eggenberger2019}. Two families of rotating stellar models are considered in the present study : models that only include transport by hydrodynamic processes (labelled as `pure rotation' in Fig. \ref{fig:figure2}) and models that include both hydrodynamic and magnetic transport processes (labelled as `Tayler instability' in Fig. \ref{fig:figure2}). The difference between these models relies on the inclusion of transport by the magnetic Tayler instability for the latter through the coefficient $\nu_{\rm TS}$ in the equation above. For both families of models, we accounted for braking of the stellar surface due to magnetised winds according to the prescription by \citet{mat15} for models of stars with an extended convective envelope like the Sun, and therefore Doris in this study. For Arthur, Barney, and Carlsberg, which are stars in the hotter part of the HR diagram (see Fig. \ref{fig:hr}) characterized by shallow convective envelopes, we assumed an inefficient braking that is inefficient enough such that it can be modelled by simply neglecting the corresponding term. We discussed the relevance of this assumption in Sec. \ref{sec:surface_rotation}. The initial values of the rotation period on the ZAMS are 0.9, 18, 17 and 9 days, for Doris, Arthur, Barney and Carlsberg, respectively. We note that using other physical prescriptions for the magnetic instability \citep[e.g.][]{Fuller2019} or for the magnetic braking \citep[e.g.][]{Garraffo2018} does not affect the conclusions of this study. In addition, the rotation period of the \citet{Benomar2015} sample is in the range $\sim$ 3 to 18 days, which is consistent with gyrochronologic surveys where the rotation period of stars of $\sim 6000K$ is mainly observed in the range $\sim$ 5 to 20 days \citep{McQuillan2014,vanSaders2019}. Hence, the rotation periods of Arthur, Barney, and Carlsberg are typical of that of stars in that temperature range. In that regard, Carlsberg has a rotation period close to the average value, while Arthur and Barney lie in the slower half.

\section{Rotational properties}
\label{sec:surface_rotation}

In the upper panels of Fig. \ref{fig:figure2}, we show the rotation profiles of Arthur and Doris by considering the two different scenarios. The rotation profiles of Barney and Carlsberg are similar to that of Arthur (see Appendix~\ref{app:supplementary_data}). The rotation behaviour is quite different between the hot targets and the cold target Doris. Figure~\ref{fig:figure2} indeed shows a higher degree of radial differential rotation for Doris than for other targets as a direct consequence of the efficient braking of the stellar surface by magnetized winds for this colder star. In the case of Doris, this results in a clear difference in the rotation profile predicted for the model with only hydrodynamic transport processes (green line in the top left panel of Fig.~\ref{fig:figure2}) compared to the one predicted for the model with magnetic instabilities (red line in the same panel of Fig.~\ref{fig:figure2}). The efficient magnetic AM transport predicts an almost flat rotation profile for Doris, except in the central layers where strong chemical gradients reduce the efficiency of this transport. AM transport by meridional circulation and shear instability is much less efficient and is not able to counteract the creation of radial differential rotation in the radiative zone leading to a core that rotates more than ten times faster than the surface. The situation is different in hotter stars as shown by the models of Arthur in the top center panel of Fig. \ref{fig:figure2}. For these models, radial differential rotation results solely from the slight contraction of the central layers and the increase of the radius, which leads to a low degree of differential rotation in the radiative interior even for models with only hydrodynamic transport processes (see top center panel of Fig.~\ref{fig:figure2}). Owing to the very efficient AM transport by the magnetic Tayler instability, models with the Tayler-Spruit dynamo predict an even flatter rotation profile than the ones with only meridional circulation and the shear instability.

As proposed by OB15, a measurement of the surface rotation rate can be compared to the asteroseismic determination of the mean internal rotation rate of the star as probed by p-modes to shed some light on the radial differential rotation and hence on AM transport in these stars. For slow rotators, the effect of the rotation can be treated as a small perturbation of the pulsation frequency. Assuming spherically symmetric rotation, the splittings simplify to \citep{Ledoux1951,Schou1994}:
\begin{align}
\delta\omega_{n,l,m} = m\beta_{n,l}\int_0^R K_{n,l}(r)\Omega(r)dr,
\end{align}
where $m$ is the azimuthal order, $\beta_{n,l}$ is a constant that depends on the radial order $n$ and on the harmonic degree $l$, and $K_{n,l}$ is the rotation kernel. For a given $(n,l)$ pair, we define the mean asteroseismic rotation rate as
\begin{align}
\label{eq:omega_surf_sismo}
\Omega_{\mathrm{sismo}}^{n,l} \equiv \int_0^R K_{n,l}(r)\Omega(r)dr.
\end{align}
We only considered the $l=1$ rotational splittings, and verified that they were consistent with the $l=2$ rotational splittings.

In the lower panels of Fig. \ref{fig:figure2}, we show the ratio between the asteroseismic rotation rate and the surface rotation rate $\Omega_{\mathrm{surf}}$ for the different models of Arthur and Doris. The open circles correspond to the theoretical values of $\Omega_{\mathrm{sismo}}^{n,l}$ computed with Eq. \eqref{eq:omega_surf_sismo}, for the different AM transport prescriptions considered in this study. The blue area and the error bars correspond to actual measurements, to highlight the detectability of the model differences. The error bars correspond to the precision of the average observational rotational splitting measured by \citet{Benomar2018} (hereafter OB18), which is mostly dependent on the signal-to-noise ratio of the modes and of the mode blending width at $\nu_{max}$, and the blue area corresponds to a solid-body rotation profile with the precision of the surface rotation rate from starspots. A value of $\Omega_{\mathrm{sismo}}^{n,l}$ compatible with the blue area means that asteroseismology do not detect a significant degree of radial differential rotation, as expected from OB15 and MN17. Models with AM transport by magnetic instabilities (red circles) are always compatible with the observations of a similar rotation rate derived from asteroseismic splittings and from an independent measurement of the surface rotation rate as reported by OB15 and MN17 for stars in the cold part of the HR diagram (Doris) as well as on the hotter side (Arthur). The efficient AM transport predicted by these models is thus in agreement with the asteroseismic constraints on the internal rotation currently available for various solar-type main-sequence stars similarly to what is found in the case of the Sun \citep[see][]{Eggenberger2019,Eggenberger2022}. 

In the case of models with only hydrodynamic AM transport, the situation is different. For targets in the cold side of the HR diagram (Doris), these models predict asteroseismic rotation rates significantly larger than surface rotation rates due to both the efficient braking of the surface by magnetized winds and the low efficiency of AM transport by meridional circulation and shear instability, which leads to a high degree of radial differential rotation in the radiative interior, in particular close to the base of the convective envelope that can be probed by rotational kernels. These rotating models with transport by only hydrodynamic processes can then be easily rejected with asteroseismic measurements of solar-type stars in the cold side of the HR diagram although the observational uncertainties are large (green circles in Fig. \ref{fig:figure2}). For hotter main-sequence stars, the difference between asteroseismic and surface rotation rates expected for models with only hydrodynamic transport is much lower due to the inefficient surface braking associated with stars with shallow convective envelopes. With the assumption of an inefficient surface braking, the radial differential rotation predicted for these models is too small in the region probed by seismology and falls within the observational uncertainty of the surface rotation rates deduced from starspots (green circles in the bottom center panel of Fig. \ref{fig:figure2}). Owing to the uncertainties in the knowledge of surface magnetic braking for stars hotter than the Sun, it is difficult to determine exactly above which effective temperature this braking is really inefficient. Stars similar to Arthur are indeed expected to be close to such a transition, but it's not absolutely clear whether the assumption of inefficient braking adopted here is fully justified. We thus investigated the impact of introducing a much more efficient braking for stars in the blue part of the HR diagram on the conclusion about internal AM transport obtained for Arthur. We computed a new model for Arthur by using the same prescription for an efficient surface braking by magnetized winds as introduced for the cooler star Doris. This model (labelled as `efficient braking' in Fig. \ref{fig:figure2}) is computed with an initial rotation period on the ZAMS of 1.6 days and shows then a higher degree of radial differential rotation as a result of this strong braking of the surface and low AM transport efficiency by hydrodynamic processes alone, which can then be discarded using asteroseismic and surface rotation rates observations similarly to the result obtained for the cooler star Doris. For Arthur, we also tested another extreme case, for which we assumed local AM conservation (labelled as `local conservation' in Fig. \ref{fig:figure2}). This model behaves similarly to the model including hydrodynamic processes alone. We thus observe that for stars in the blue part of the HR diagram, it is difficult to reject rotating models with an inefficient of AM transport based on combined asteroseismic and surface rotation rate measurements, as only cases with an efficient surface braking by magnetized winds could be detected.


\section{Conclusions}
\label{sec:conclusions}
We carried out a detailed modelling of four \emph{Kepler LEGACY} targets, three laying in the hotter side of the HR diagram and with different evolutionary stages in the MS, and one in the colder side. In Sec. \ref{sec:modelling_and_rotation_profiles}, we describe the asteroseismic modelling procedure together with the computation of rotating models with only hydrodynamic AM transport and models with both hydrodynamic and magnetic transport. The asteroseismic rotation rates were then computed for these different models and compared with the surface rotation rates deduced from observations of starspots in Sec. \ref{sec:surface_rotation}.

For main-sequence stars in the hotter part of the HR diagram (masses typically above $\sim 1.2M_\odot$ at solar metallicity), models with only hydrodynamic transport processes and models with additional transport by magnetic instabilities are found to be consistent with measurements reported by OB15 and MN17 who observed a low degree of radial differential rotation between the radiative and convective zones. For these stars, which constitute a significant fraction of the \emph{Kepler LEGACY} sample, combining asteroseismic constraints from splittings of pressure modes and surface rotation rates does not allow to conclude on the need for an efficient AM transport in addition to the sole transport by meridional circulation and shear instability. Even the model assuming local AM conservation cannot be ruled out. This is because rotational kernels probe a region close below the BCZ, where radial differential rotation can be low for these stars with shallow convective envelopes. If an unlikely efficient surface braking is assumed for these hotter stars, the degree of radial differential rotation would be incompatible with the observations. Further investigations on that specific point, beyond the scope of this study, are required to find if this signature appears in the available observational data or not. In the colder part of the HR diagram, the situation is different due to the efficient braking of the stellar surface by magnetised winds. We observed a clear disagreement between the rotational properties of models including only hydrodynamic processes and asteroseismic constraints, while models with magnetic fields correctly reproduce the observations, similarly to the solar case. This disagreement allows to easily rule out models with an inefficient transport of AM in that part of the HR diagram.


\section*{Acknowledgements}
J.B. and G.B. acknowledge funding from the SNF AMBIZIONE grant No 185805 (Seismic inversions and modelling of transport processes in stars). P.E. has received funding from the European Research Council (ERC) under the European Union’s Horizon 2020 research and innovation programme (grant agreement No. 833925, project STAREX).


\bibliography{bibliography.bib}

\begin{thebibliography}{70}
\expandafter\ifx\csname natexlab\endcsname\relax\def\natexlab#1{#1}\fi

\bibitem[{{Baglin} {et~al.}(2009){Baglin}, {Auvergne}, {Barge}, {Deleuil},
  {Michel}, \& {CoRoT Exoplanet Science Team}}]{Baglin2009}
{Baglin}, A., {Auvergne}, M., {Barge}, P., {et~al.} 2009, in IAU Symposium,
  Vol. 253, Transiting Planets, ed. F.~{Pont}, D.~{Sasselov}, \& M.~J.
  {Holman}, 71--81

\bibitem[{{Bailer-Jones} {et~al.}(2021){Bailer-Jones}, {Rybizki}, {Fouesneau},
  {Demleitner}, \& {Andrae}}]{Bailer-Jones2021}
{Bailer-Jones}, C.~A.~L., {Rybizki}, J., {Fouesneau}, M., {Demleitner}, M., \&
  {Andrae}, R. 2021, \aj, 161, 147

\bibitem[{{Beck} {et~al.}(2012){Beck}, {Montalban}, {Kallinger}, {De Ridder},
  {Aerts}, {Garc{\'\i}a}, {Hekker}, {Dupret}, {Mosser}, {Eggenberger},
  {Stello}, {Elsworth}, {Frandsen}, {Carrier}, {Hillen}, {Gruberbauer},
  {Christensen-Dalsgaard}, {Miglio}, {Valentini}, {Bedding}, {Kjeldsen},
  {Girouard}, {Hall}, \& {Ibrahim}}]{Beck2012}
{Beck}, P.~G., {Montalban}, J., {Kallinger}, T., {et~al.} 2012, \nat, 481, 55

\bibitem[{{Benomar} {et~al.}(2018){Benomar}, {Bazot}, {Nielsen}, {Gizon},
  {Sekii}, {Takata}, {Hotta}, {Hanasoge}, {Sreenivasan}, \&
  {Christensen-Dalsgaard}}]{Benomar2018}
{Benomar}, O., {Bazot}, M., {Nielsen}, M.~B., {et~al.} 2018, Science, 361, 1231

\bibitem[{{Benomar} {et~al.}(2015){Benomar}, {Takata}, {Shibahashi},
  {Ceillier}, \& {Garc{\'\i}a}}]{Benomar2015}
{Benomar}, O., {Takata}, M., {Shibahashi}, H., {Ceillier}, T., \&
  {Garc{\'\i}a}, R.~A. 2015, \mnras, 452, 2654

\bibitem[{{B{\'e}trisey} {et~al.}(2022){B{\'e}trisey}, {Pezzotti}, {Buldgen},
  {Khan}, {Eggenberger}, {Salmon}, \& {Miglio}}]{Betrisey2022}
{B{\'e}trisey}, J., {Pezzotti}, C., {Buldgen}, G., {et~al.} 2022, \aap, 659,
  A56

\bibitem[{{Borucki} {et~al.}(2010){Borucki}, {Koch}, {Basri}, {Batalha},
  {Brown}, {Caldwell}, {Caldwell}, {Christensen-Dalsgaard}, {Cochran},
  {DeVore}, {Dunham}, {Dupree}, {Gautier}, {Geary}, {Gilliland}, {Gould},
  {Howell}, {Jenkins}, {Kondo}, {Latham}, {Marcy}, {Meibom}, {Kjeldsen},
  {Lissauer}, {Monet}, {Morrison}, {Sasselov}, {Tarter}, {Boss}, {Brownlee},
  {Owen}, {Buzasi}, {Charbonneau}, {Doyle}, {Fortney}, {Ford}, {Holman},
  {Seager}, {Steffen}, {Welsh}, {Rowe}, {Anderson}, {Buchhave}, {Ciardi},
  {Walkowicz}, {Sherry}, {Horch}, {Isaacson}, {Everett}, {Fischer}, {Torres},
  {Johnson}, {Endl}, {MacQueen}, {Bryson}, {Dotson}, {Haas}, {Kolodziejczak},
  {Van Cleve}, {Chandrasekaran}, {Twicken}, {Quintana}, {Clarke}, {Allen},
  {Li}, {Wu}, {Tenenbaum}, {Verner}, {Bruhweiler}, {Barnes}, \&
  {Prsa}}]{Borucki2010}
{Borucki}, W.~J., {Koch}, D., {Basri}, G., {et~al.} 2010, Science, 327, 977

\bibitem[{{Buldgen} {et~al.}(2022){Buldgen}, {B{\'e}trisey}, {Roxburgh},
  {Vorontsov}, \& {Reese}}]{Buldgen2022_Fr}
{Buldgen}, G., {B{\'e}trisey}, J., {Roxburgh}, I.~W., {Vorontsov}, S.~V., \&
  {Reese}, D.~R. 2022, Frontiers in Astronomy and Space Sciences, 9, 942373

\bibitem[{{Buldgen} {et~al.}(2019){Buldgen}, {Farnir}, {Pezzotti},
  {Eggenberger}, {Salmon}, {Montalban}, {Ferguson}, {Khan}, {Bourrier},
  {Rendle}, {Meynet}, {Miglio}, \& {Noels}}]{Buldgen2019f}
{Buldgen}, G., {Farnir}, M., {Pezzotti}, C., {et~al.} 2019, \aap, 630, A126

\bibitem[{{Ceillier} {et~al.}(2013){Ceillier}, {Eggenberger}, {Garc{\'{\i}}a},
  \& {Mathis}}]{Ceillier2013}
{Ceillier}, T., {Eggenberger}, P., {Garc{\'{\i}}a}, R.~A., \& {Mathis}, S.
  2013, \aap, 555, A54

\bibitem[{{Chaboyer} {et~al.}(1995){Chaboyer}, {Demarque}, \&
  {Pinsonneault}}]{Chaboyer1995}
{Chaboyer}, B., {Demarque}, P., \& {Pinsonneault}, M.~H. 1995, \apj, 441, 865

\bibitem[{{Charbonneau} \& {MacGregor}(1993)}]{Charbonneau1993}
{Charbonneau}, P. \& {MacGregor}, K.~B. 1993, \apj, 417, 762

\bibitem[{{Charbonnel} \& {Talon}(2005)}]{Charbonnel&Talon2005}
{Charbonnel}, C. \& {Talon}, S. 2005, Science, 309, 2189

\bibitem[{{Deheuvels} {et~al.}(2015){Deheuvels}, {Ballot}, {Beck}, {Mosser},
  {{\O}stensen}, {Garc{\'\i}a}, \& {Goupil}}]{Deheuvels2015}
{Deheuvels}, S., {Ballot}, J., {Beck}, P.~G., {et~al.} 2015, \aap, 580, A96

\bibitem[{{Deheuvels} {et~al.}(2014){Deheuvels}, {Do{\u{g}}an}, {Goupil},
  {Appourchaux}, {Benomar}, {Bruntt}, {Campante}, {Casagrande}, {Ceillier},
  {Davies}, {De Cat}, {Fu}, {Garc{\'\i}a}, {Lobel}, {Mosser}, {Reese},
  {Regulo}, {Schou}, {Stahn}, {Thygesen}, {Yang}, {Chaplin},
  {Christensen-Dalsgaard}, {Eggenberger}, {Gizon}, {Mathis},
  {Molenda-{\.Z}akowicz}, \& {Pinsonneault}}]{Deheuvels2014}
{Deheuvels}, S., {Do{\u{g}}an}, G., {Goupil}, M.~J., {et~al.} 2014, \aap, 564,
  A27

\bibitem[{{Deheuvels} {et~al.}(2012){Deheuvels}, {Garc{\'\i}a}, {Chaplin},
  {Basu}, {Antia}, {Appourchaux}, {Benomar}, {Davies}, {Elsworth}, {Gizon},
  {Goupil}, {Reese}, {Regulo}, {Schou}, {Stahn}, {Casagrande},
  {Christensen-Dalsgaard}, {Fischer}, {Hekker}, {Kjeldsen}, {Mathur}, {Mosser},
  {Pinsonneault}, {Valenti}, {Christiansen}, {Kinemuchi}, \&
  {Mullally}}]{Deheuvels2012}
{Deheuvels}, S., {Garc{\'\i}a}, R.~A., {Chaplin}, W.~J., {et~al.} 2012, \apj,
  756, 19

\bibitem[{{Deheuvels} {et~al.}(2017){Deheuvels}, {Ouazzani}, \&
  {Basu}}]{Deheuvels2017}
{Deheuvels}, S., {Ouazzani}, R.~M., \& {Basu}, S. 2017, \aap, 605, A75

\bibitem[{{Di Mauro} {et~al.}(2016){Di Mauro}, {Ventura}, {Cardini}, {Stello},
  {Christensen-Dalsgaard}, {Dziembowski}, {Patern{\`o}}, {Beck}, {Bloemen},
  {Davies}, {De Smedt}, {Elsworth}, {Garc{\'\i}a}, {Hekker}, {Mosser}, \&
  {Tkachenko}}]{DiMauro2016}
{Di Mauro}, M.~P., {Ventura}, R., {Cardini}, D., {et~al.} 2016, \apj, 817, 65

\bibitem[{{Di Mauro} {et~al.}(2018){Di Mauro}, {Ventura}, {Corsaro}, \&
  {Lustosa De Moura}}]{Dimauro2018}
{Di Mauro}, M.~P., {Ventura}, R., {Corsaro}, E., \& {Lustosa De Moura}, B.
  2018, \apj, 862, 9

\bibitem[{{Eff-Darwich} \& {Korzennik}(2013)}]{Eff-Darwich&Korzennik2013}
{Eff-Darwich}, A. \& {Korzennik}, S.~G. 2013, \solphys, 287, 43

\bibitem[{{Eggenberger} {et~al.}(2022){Eggenberger}, {Buldgen}, {Salmon},
  {Noels}, {Grevesse}, \& {Asplund}}]{Eggenberger2022}
{Eggenberger}, P., {Buldgen}, G., {Salmon}, S.~J.~A.~J., {et~al.} 2022, Nature
  Astronomy, 6, 788

\bibitem[{{Eggenberger} {et~al.}(2019){Eggenberger}, {Deheuvels}, {Miglio},
  {Ekstr{\"o}m}, {Georgy}, {Meynet}, {Lagarde}, {Salmon}, {Buldgen},
  {Montalb{\'a}n}, {Spada}, \& {Ballot}}]{Eggenberger2019}
{Eggenberger}, P., {Deheuvels}, S., {Miglio}, A., {et~al.} 2019, \aap, 621, A66

\bibitem[{{Eggenberger} {et~al.}(2017){Eggenberger}, {Lagarde}, {Miglio},
  {Montalb{\'a}n}, {Ekstr{\"o}m}, {Georgy}, {Meynet}, {Salmon}, {Ceillier},
  {Garc{\'{\i}}a}, {Mathis}, {Deheuvels}, {Maeder}, {den Hartogh}, \&
  {Hirschi}}]{Eggenberger2017}
{Eggenberger}, P., {Lagarde}, N., {Miglio}, A., {et~al.} 2017, \aap, 599, A18

\bibitem[{{Eggenberger} {et~al.}(2005){Eggenberger}, {Maeder}, \&
  {Meynet}}]{Eggenberger2005mag}
{Eggenberger}, P., {Maeder}, A., \& {Meynet}, G. 2005, \aap, 440, L9

\bibitem[{{Eggenberger} {et~al.}(2008){Eggenberger}, {Meynet}, {Maeder},
  {Hirschi}, {Charbonnel}, {Talon}, \& {Ekstr{\"o}m}}]{Eggenberger2008}
{Eggenberger}, P., {Meynet}, G., {Maeder}, A., {et~al.} 2008, \apss, 316, 43

\bibitem[{{Eggenberger} {et~al.}(2012){Eggenberger}, {Montalb{\'a}n}, \&
  {Miglio}}]{Eggenberger2012rg}
{Eggenberger}, P., {Montalb{\'a}n}, J., \& {Miglio}, A. 2012, \aap, 544, L4

\bibitem[{{Farnir} {et~al.}(2020){Farnir}, {Dupret}, {Buldgen}, {Salmon},
  {Noels}, {Pin{\c{c}}on}, {Pezzotti}, \& {Eggenberger}}]{Farnir2020}
{Farnir}, M., {Dupret}, M.~A., {Buldgen}, G., {et~al.} 2020, \aap, 644, A37

\bibitem[{{Fellay} {et~al.}(2021){Fellay}, {Buldgen}, {Eggenberger}, {Khan},
  {Salmon}, {Miglio}, \& {Montalb{\'a}n}}]{Fellay2021}
{Fellay}, L., {Buldgen}, G., {Eggenberger}, P., {et~al.} 2021, \aap, 654, A133

\bibitem[{{Fuller} {et~al.}(2019){Fuller}, {Piro}, \& {Jermyn}}]{Fuller2019}
{Fuller}, J., {Piro}, A.~L., \& {Jermyn}, A.~S. 2019, \mnras, 485, 3661

\bibitem[{{Furlan} {et~al.}(2018){Furlan}, {Ciardi}, {Cochran}, {Everett},
  {Latham}, {Marcy}, {Buchhave}, {Endl}, {Isaacson}, {Petigura}, {Gautier},
  {Huber}, {Bieryla}, {Borucki}, {Brugamyer}, {Caldwell}, {Cochran}, {Howard},
  {Howell}, {Johnson}, {MacQueen}, {Quinn}, {Robertson}, {Mathur}, \&
  {Batalha}}]{Furlan2018}
{Furlan}, E., {Ciardi}, D.~R., {Cochran}, W.~D., {et~al.} 2018, \apj, 861, 149

\bibitem[{{Gaia Collaboration} {et~al.}(2021){Gaia Collaboration}, {Brown},
  {Vallenari}, {Prusti}, {de Bruijne}, {Babusiaux}, {Biermann}, {Creevey},
  {Evans}, {Eyer}, {Hutton}, {Jansen}, {Jordi}, {Klioner}, {Lammers},
  {Lindegren}, {Luri}, {Mignard}, {Panem}, {Pourbaix}, {Randich}, {Sartoretti},
  {Soubiran}, {Walton}, {Arenou}, {Bailer-Jones}, {Bastian}, {Cropper},
  {Drimmel}, {Katz}, {Lattanzi}, {van Leeuwen}, {Bakker}, {Cacciari},
  {Casta{\~n}eda}, {De Angeli}, {Ducourant}, {Fabricius}, {Fouesneau},
  {Fr{\'e}mat}, {Guerra}, {Guerrier}, {Guiraud}, {Jean-Antoine Piccolo},
  {Masana}, {Messineo}, {Mowlavi}, {Nicolas}, {Nienartowicz}, {Pailler},
  {Panuzzo}, {Riclet}, {Roux}, {Seabroke}, {Sordo}, {Tanga}, {Th{\'e}venin},
  {Gracia-Abril}, {Portell}, {Teyssier}, {Altmann}, {Andrae}, {Bellas-Velidis},
  {Benson}, {Berthier}, {Blomme}, {Brugaletta}, {Burgess}, {Busso}, {Carry},
  {Cellino}, {Cheek}, {Clementini}, {Damerdji}, {Davidson}, {Delchambre},
  {Dell'Oro}, {Fern{\'a}ndez-Hern{\'a}ndez}, {Galluccio}, {Garc{\'\i}a-Lario},
  {Garcia-Reinaldos}, {Gonz{\'a}lez-N{\'u}{\~n}ez}, {Gosset}, {Haigron},
  {Halbwachs}, {Hambly}, {Harrison}, {Hatzidimitriou}, {Heiter},
  {Hern{\'a}ndez}, {Hestroffer}, {Hodgkin}, {Holl}, {Jan{\ss}en}, {Jevardat de
  Fombelle}, {Jordan}, {Krone-Martins}, {Lanzafame}, {L{\"o}ffler}, {Lorca},
  {Manteiga}, {Marchal}, {Marrese}, {Moitinho}, {Mora}, {Muinonen}, {Osborne},
  {Pancino}, {Pauwels}, {Petit}, {Recio-Blanco}, {Richards}, {Riello},
  {Rimoldini}, {Robin}, {Roegiers}, {Rybizki}, {Sarro}, {Siopis}, {Smith},
  {Sozzetti}, {Ulla}, {Utrilla}, {van Leeuwen}, {van Reeven}, {Abbas}, {Abreu
  Aramburu}, {Accart}, {Aerts}, {Aguado}, {Ajaj}, {Altavilla}, {{\'A}lvarez},
  {{\'A}lvarez Cid-Fuentes}, {Alves}, {Anderson}, {Anglada Varela}, {Antoja},
  {Audard}, {Baines}, {Baker}, {Balaguer-N{\'u}{\~n}ez}, {Balbinot}, {Balog},
  {Barache}, {Barbato}, {Barros}, {Barstow}, {Bartolom{\'e}}, {Bassilana},
  {Bauchet}, {Baudesson-Stella}, {Becciani}, {Bellazzini}, {Bernet}, {Bertone},
  {Bianchi}, {Blanco-Cuaresma}, {Boch}, {Bombrun}, {Bossini}, {Bouquillon},
  {Bragaglia}, {Bramante}, {Breedt}, {Bressan}, {Brouillet}, {Bucciarelli},
  {Burlacu}, {Busonero}, {Butkevich}, {Buzzi}, {Caffau}, {Cancelliere},
  {C{\'a}novas}, {Cantat-Gaudin}, {Carballo}, {Carlucci}, {Carnerero},
  {Carrasco}, {Casamiquela}, {Castellani}, {Castro-Ginard}, {Castro Sampol},
  {Chaoul}, {Charlot}, {Chemin}, {Chiavassa}, {Cioni}, {Comoretto}, {Cooper},
  {Cornez}, {Cowell}, {Crifo}, {Crosta}, {Crowley}, {Dafonte}, {Dapergolas},
  {David}, {David}, {de Laverny}, {De Luise}, {De March}, {De Ridder}, {de
  Souza}, {de Teodoro}, {de Torres}, {del Peloso}, {del Pozo}, {Delbo},
  {Delgado}, {Delgado}, {Delisle}, {Di Matteo}, {Diakite}, {Diener},
  {Distefano}, {Dolding}, {Eappachen}, {Edvardsson}, {Enke}, {Esquej}, {Fabre},
  {Fabrizio}, {Faigler}, {Fedorets}, {Fernique}, {Fienga}, {Figueras},
  {Fouron}, {Fragkoudi}, {Fraile}, {Franke}, {Gai}, {Garabato},
  {Garcia-Gutierrez}, {Garc{\'\i}a-Torres}, {Garofalo}, {Gavras}, {Gerlach},
  {Geyer}, {Giacobbe}, {Gilmore}, {Girona}, {Giuffrida}, {Gomel}, {Gomez},
  {Gonzalez-Santamaria}, {Gonz{\'a}lez-Vidal}, {Granvik},
  {Guti{\'e}rrez-S{\'a}nchez}, {Guy}, {Hauser}, {Haywood}, {Helmi}, {Hidalgo},
  {Hilger}, {H{\l}adczuk}, {Hobbs}, {Holland}, {Huckle}, {Jasniewicz},
  {Jonker}, {Juaristi Campillo}, {Julbe}, {Karbevska}, {Kervella}, {Khanna},
  {Kochoska}, {Kontizas}, {Kordopatis}, {Korn}, {Kostrzewa-Rutkowska},
  {Kruszy{\'n}ska}, {Lambert}, {Lanza}, {Lasne}, {Le Campion}, {Le Fustec},
  {Lebreton}, {Lebzelter}, {Leccia}, {Leclerc}, {Lecoeur-Taibi}, {Liao},
  {Licata}, {Lindstr{\o}m}, {Lister}, {Livanou}, {Lobel}, {Madrero Pardo},
  {Managau}, {Mann}, {Marchant}, {Marconi}, {Marcos Santos}, {Marinoni},
  {Marocco}, {Marshall}, {Martin Polo}, {Mart{\'\i}n-Fleitas}, {Masip},
  {Massari}, {Mastrobuono-Battisti}, {Mazeh}, {McMillan}, {Messina},
  {Michalik}, {Millar}, {Mints}, {Molina}, {Molinaro}, {Moln{\'a}r},
  {Montegriffo}, {Mor}, {Morbidelli}, {Morel}, {Morris}, {Mulone}, {Munoz},
  {Muraveva}, {Murphy}, {Musella}, {Noval}, {Ord{\'e}novic}, {Orr{\`u}},
  {Osinde}, {Pagani}, {Pagano}, {Palaversa}, {Palicio}, {Panahi}, {Pawlak},
  {Pe{\~n}alosa Esteller}, {Penttil{\"a}}, {Piersimoni}, {Pineau}, {Plachy},
  {Plum}, {Poggio}, {Poretti}, {Poujoulet}, {Pr{\v{s}}a}, {Pulone}, {Racero},
  {Ragaini}, {Rainer}, {Raiteri}, {Rambaux}, {Ramos}, {Ramos-Lerate}, {Re
  Fiorentin}, {Regibo}, {Reyl{\'e}}, {Ripepi}, {Riva}, {Rixon}, {Robichon},
  {Robin}, {Roelens}, {Rohrbasser}, {Romero-G{\'o}mez}, {Rowell}, {Royer},
  {Rybicki}, {Sadowski}, {Sagrist{\`a} Sell{\'e}s}, {Sahlmann}, {Salgado},
  {Salguero}, {Samaras}, {Sanchez Gimenez}, {Sanna}, {Santove{\~n}a},
  {Sarasso}, {Schultheis}, {Sciacca}, {Segol}, {Segovia}, {S{\'e}gransan},
  {Semeux}, {Shahaf}, {Siddiqui}, {Siebert}, {Siltala}, {Slezak}, {Smart},
  {Solano}, {Solitro}, {Souami}, {Souchay}, {Spagna}, {Spoto}, {Steele},
  {Steidelm{\"u}ller}, {Stephenson}, {S{\"u}veges}, {Szabados}, {Szegedi-Elek},
  {Taris}, {Tauran}, {Taylor}, {Teixeira}, {Thuillot}, {Tonello}, {Torra},
  {Torra}, {Turon}, {Unger}, {Vaillant}, {van Dillen}, {Vanel}, {Vecchiato},
  {Viala}, {Vicente}, {Voutsinas}, {Weiler}, {Wevers}, {Wyrzykowski}, {Yoldas},
  {Yvard}, {Zhao}, {Zorec}, {Zucker}, {Zurbach}, \& {Zwitter}}]{Gaia2021}
{Gaia Collaboration}, {Brown}, A.~G.~A., {Vallenari}, A., {et~al.} 2021, \aap,
  649, A1

\bibitem[{{Garc{\'\i}a} {et~al.}(2014){Garc{\'\i}a}, {Ceillier}, {Salabert},
  {Mathur}, {van Saders}, {Pinsonneault}, {Ballot}, {Beck}, {Bloemen},
  {Campante}, {Davies}, {do Nascimento}, {Mathis}, {Metcalfe}, {Nielsen},
  {Su{\'a}rez}, {Chaplin}, {Jim{\'e}nez}, \& {Karoff}}]{Garcia2014}
{Garc{\'\i}a}, R.~A., {Ceillier}, T., {Salabert}, D., {et~al.} 2014, \aap, 572,
  A34

\bibitem[{{Garraffo} {et~al.}(2018){Garraffo}, {Drake}, {Dotter}, {Choi},
  {Burke}, {Moschou}, {Alvarado-G{\'o}mez}, {Kashyap}, \&
  {Cohen}}]{Garraffo2018}
{Garraffo}, C., {Drake}, J.~J., {Dotter}, A., {et~al.} 2018, \apj, 862, 90

\bibitem[{{Gehan} {et~al.}(2018){Gehan}, {Mosser}, {Michel}, {Samadi}, \&
  {Kallinger}}]{Gehan2018}
{Gehan}, C., {Mosser}, B., {Michel}, E., {Samadi}, R., \& {Kallinger}, T. 2018,
  \aap, 616, A24

\bibitem[{{Gough} \& {McIntyre}(1998)}]{Gough&McIntyre1998}
{Gough}, D.~O. \& {McIntyre}, M.~E. 1998, \nat, 394, 755

\bibitem[{{Kraft}(1967)}]{Kraft1967}
{Kraft}, R.~P. 1967, \apj, 150, 551

\bibitem[{{Kurtz} {et~al.}(2014){Kurtz}, {Saio}, {Takata}, {Shibahashi},
  {Murphy}, \& {Sekii}}]{Kurtz2014}
{Kurtz}, D.~W., {Saio}, H., {Takata}, M., {et~al.} 2014, \mnras, 444, 102

\bibitem[{{Ledoux}(1951)}]{Ledoux1951}
{Ledoux}, P. 1951, \apj, 114, 373

\bibitem[{{Li} {et~al.}(2020){Li}, {Van Reeth}, {Bedding}, {Murphy}, {Antoci},
  {Ouazzani}, \& {Barbara}}]{Li2020}
{Li}, G., {Van Reeth}, T., {Bedding}, T.~R., {et~al.} 2020, \mnras, 491, 3586

\bibitem[{{Lund} {et~al.}(2017){Lund}, {Silva Aguirre}, {Davies}, {Chaplin},
  {Christensen-Dalsgaard}, {Houdek}, {White}, {Bedding}, {Ball}, {Huber},
  {Antia}, {Lebreton}, {Latham}, {Handberg}, {Verma}, {Basu}, {Casagrande},
  {Justesen}, {Kjeldsen}, \& {Mosumgaard}}]{Lund2017}
{Lund}, M.~N., {Silva Aguirre}, V., {Davies}, G.~R., {et~al.} 2017, \apj, 835,
  172

\bibitem[{{Marques} {et~al.}(2013){Marques}, {Goupil}, {Lebreton}, {Talon},
  {Palacios}, {Belkacem}, {Ouazzani}, {Mosser}, {Moya}, {Morel}, {Pichon},
  {Mathis}, {Zahn}, {Turck-Chi{\`e}ze}, \& {Nghiem}}]{Marques2013}
{Marques}, J.~P., {Goupil}, M.~J., {Lebreton}, Y., {et~al.} 2013, \aap, 549,
  A74

\bibitem[{{Matt} {et~al.}(2015){Matt}, {Brun}, {Baraffe}, {Bouvier}, \&
  {Chabrier}}]{mat15}
{Matt}, S.~P., {Brun}, A.~S., {Baraffe}, I., {Bouvier}, J., \& {Chabrier}, G.
  2015, \apjl, 799, L23

\bibitem[{{McQuillan} {et~al.}(2014){McQuillan}, {Mazeh}, \&
  {Aigrain}}]{McQuillan2014}
{McQuillan}, A., {Mazeh}, T., \& {Aigrain}, S. 2014, \apjs, 211, 24

\bibitem[{{Mestel} \& {Weiss}(1987)}]{Mestel&Weiss1987}
{Mestel}, L. \& {Weiss}, N.~O. 1987, \mnras, 226, 123

\bibitem[{{Mosser} {et~al.}(2012){Mosser}, {Goupil}, {Belkacem}, {Marques},
  {Beck}, {Bloemen}, {De Ridder}, {Barban}, {Deheuvels}, {Elsworth}, {Hekker},
  {Kallinger}, {Ouazzani}, {Pinsonneault}, {Samadi}, {Stello}, {Garc{\'\i}a},
  {Klaus}, {Li}, {Mathur}, \& {Morris}}]{Mosser2012}
{Mosser}, B., {Goupil}, M.~J., {Belkacem}, K., {et~al.} 2012, \aap, 548, A10

\bibitem[{{Moyano} {et~al.}(2022){Moyano}, {Eggenberger}, {Meynet}, {Gehan},
  {Mosser}, {Buldgen}, \& {Salmon}}]{Moyano2022}
{Moyano}, F.~D., {Eggenberger}, P., {Meynet}, G., {et~al.} 2022, \aap, 663,
  A180

\bibitem[{{Murphy} {et~al.}(2016){Murphy}, {Fossati}, {Bedding}, {Saio},
  {Kurtz}, {Grassitelli}, \& {Wang}}]{Murphy2016}
{Murphy}, S.~J., {Fossati}, L., {Bedding}, T.~R., {et~al.} 2016, \mnras, 459,
  1201

\bibitem[{{Nielsen} {et~al.}(2013){Nielsen}, {Gizon}, {Schunker}, \&
  {Karoff}}]{Nielsen2013}
{Nielsen}, M.~B., {Gizon}, L., {Schunker}, H., \& {Karoff}, C. 2013, \aap, 557,
  L10

\bibitem[{{Nielsen} {et~al.}(2017){Nielsen}, {Schunker}, {Gizon}, {Schou}, \&
  {Ball}}]{Nielsen2017}
{Nielsen}, M.~B., {Schunker}, H., {Gizon}, L., {Schou}, J., \& {Ball}, W.~H.
  2017, \aap, 603, A6

\bibitem[{{Ouazzani} {et~al.}(2019){Ouazzani}, {Marques}, {Goupil},
  {Christophe}, {Antoci}, {Salmon}, \& {Ballot}}]{Ouazzani2019}
{Ouazzani}, R.~M., {Marques}, J.~P., {Goupil}, M.~J., {et~al.} 2019, \aap, 626,
  A121

\bibitem[{{Pinsonneault} {et~al.}(1989){Pinsonneault}, {Kawaler}, {Sofia}, \&
  {Demarque}}]{Pinsonneault1989}
{Pinsonneault}, M.~H., {Kawaler}, S.~D., {Sofia}, S., \& {Demarque}, P. 1989,
  \apj, 338, 424

\bibitem[{{Reese} {et~al.}(2012){Reese}, {Marques}, {Goupil}, {Thompson}, \&
  {Deheuvels}}]{Reese2012}
{Reese}, D.~R., {Marques}, J.~P., {Goupil}, M.~J., {Thompson}, M.~J., \&
  {Deheuvels}, S. 2012, \aap, 539, A63

\bibitem[{{Reinhold} {et~al.}(2013){Reinhold}, {Reiners}, \&
  {Basri}}]{Reinhold2013}
{Reinhold}, T., {Reiners}, A., \& {Basri}, G. 2013, \aap, 560, A4

\bibitem[{{Rendle} {et~al.}(2019){Rendle}, {Buldgen}, {Miglio}, {Reese},
  {Noels}, {Davies}, {Campante}, {Chaplin}, {Lund}, {Kuszlewicz}, {Scott},
  {Scuflaire}, {Ball}, {Smetana}, \& {Nsamba}}]{Rendle2019}
{Rendle}, B.~M., {Buldgen}, G., {Miglio}, A., {et~al.} 2019, \mnras, 484, 771

\bibitem[{{Ricker} {et~al.}(2015){Ricker}, {Winn}, {Vanderspek}, {Latham},
  {Bakos}, {Bean}, {Berta-Thompson}, {Brown}, {Buchhave}, {Butler}, {Butler},
  {Chaplin}, {Charbonneau}, {Christensen-Dalsgaard}, {Clampin}, {Deming},
  {Doty}, {De Lee}, {Dressing}, {Dunham}, {Endl}, {Fressin}, {Ge}, {Henning},
  {Holman}, {Howard}, {Ida}, {Jenkins}, {Jernigan}, {Johnson}, {Kaltenegger},
  {Kawai}, {Kjeldsen}, {Laughlin}, {Levine}, {Lin}, {Lissauer}, {MacQueen},
  {Marcy}, {McCullough}, {Morton}, {Narita}, {Paegert}, {Palle}, {Pepe},
  {Pepper}, {Quirrenbach}, {Rinehart}, {Sasselov}, {Sato}, {Seager},
  {Sozzetti}, {Stassun}, {Sullivan}, {Szentgyorgyi}, {Torres}, {Udry}, \&
  {Villasenor}}]{Ricker2015}
{Ricker}, G.~R., {Winn}, J.~N., {Vanderspek}, R., {et~al.} 2015, Journal of
  Astronomical Telescopes, Instruments, and Systems, 1, 014003

\bibitem[{{Roxburgh}(2017)}]{Roxburgh2017}
{Roxburgh}, I.~W. 2017, \aap, 604, A42

\bibitem[{{Roxburgh} \& {Vorontsov}(2003)}]{Roxburgh&Vorontsov2003}
{Roxburgh}, I.~W. \& {Vorontsov}, S.~V. 2003, \aap, 411, 215

\bibitem[{{R\"udiger} \& {Kitchatinov}(1996)}]{Ruediger&Kitchatinov1996}
{R\"udiger}, G. \& {Kitchatinov}, L.~L. 1996, \apj, 466, 1078

\bibitem[{{Saio} {et~al.}(2015){Saio}, {Kurtz}, {Takata}, {Shibahashi},
  {Murphy}, {Sekii}, \& {Bedding}}]{Saio2015}
{Saio}, H., {Kurtz}, D.~W., {Takata}, M., {et~al.} 2015, \mnras, 447, 3264

\bibitem[{{Saio} {et~al.}(2021){Saio}, {Takata}, {Lee}, {Li}, \& {Van
  Reeth}}]{Saio2021}
{Saio}, H., {Takata}, M., {Lee}, U., {Li}, G., \& {Van Reeth}, T. 2021, \mnras,
  502, 5856

\bibitem[{{Schou} {et~al.}(1998){Schou}, {Antia}, {Basu}, {Bogart}, {Bush},
  {Chitre}, {Christensen-Dalsgaard}, {Di Mauro}, {Dziembowski}, {Eff-Darwich},
  {Gough}, {Haber}, {Hoeksema}, {Howe}, {Korzennik}, {Kosovichev}, {Larsen},
  {Pijpers}, {Scherrer}, {Sekii}, {Tarbell}, {Title}, {Thompson}, \&
  {Toomre}}]{Schou1998}
{Schou}, J., {Antia}, H.~M., {Basu}, S., {et~al.} 1998, \apj, 505, 390

\bibitem[{{Schou} {et~al.}(1994){Schou}, {Christensen-Dalsgaard}, \&
  {Thompson}}]{Schou1994}
{Schou}, J., {Christensen-Dalsgaard}, J., \& {Thompson}, M.~J. 1994, \apj, 433,
  389

\bibitem[{{Scuflaire} {et~al.}(2008{\natexlab{a}}){Scuflaire}, {Montalb{\'a}n},
  {Th{\'e}ado}, {Bourge}, {Miglio}, {Godart}, {Thoul}, \&
  {Noels}}]{Scuflaire2008a}
{Scuflaire}, R., {Montalb{\'a}n}, J., {Th{\'e}ado}, S., {et~al.}
  2008{\natexlab{a}}, \apss, 316, 149

\bibitem[{{Scuflaire} {et~al.}(2008{\natexlab{b}}){Scuflaire}, {Th{\'e}ado},
  {Montalb{\'a}n}, {Miglio}, {Bourge}, {Godart}, {Thoul}, \&
  {Noels}}]{Scuflaire2008b}
{Scuflaire}, R., {Th{\'e}ado}, S., {Montalb{\'a}n}, J., {et~al.}
  2008{\natexlab{b}}, \apss, 316, 83

\bibitem[{{Spruit}(2002)}]{Spruit2002}
{Spruit}, H.~C. 2002, \aap, 381, 923

\bibitem[{{Talon} {et~al.}(1997){Talon}, {Zahn}, {Maeder}, \& {Meynet}}]{tal97}
{Talon}, S., {Zahn}, J.-P., {Maeder}, A., \& {Meynet}, G. 1997, \aap, 322, 209

\bibitem[{{Thompson} {et~al.}(2003){Thompson}, {Christensen-Dalsgaard},
  {Miesch}, \& {Toomre}}]{Thompson2003}
{Thompson}, M.~J., {Christensen-Dalsgaard}, J., {Miesch}, M.~S., \& {Toomre},
  J. 2003, \araa, 41, 599

\bibitem[{{van Saders} {et~al.}(2019){van Saders}, {Pinsonneault}, \&
  {Barbieri}}]{vanSaders2019}
{van Saders}, J.~L., {Pinsonneault}, M.~H., \& {Barbieri}, M. 2019, \apj, 872,
  128

\bibitem[{{Zahn}(1992)}]{Zahn1992}
{Zahn}, J.~P. 1992, \aap, 265, 115

\bibitem[{{Zahn} {et~al.}(1997){Zahn}, {Talon}, \& {Matias}}]{Zahn1997}
{Zahn}, J.-P., {Talon}, S., \& {Matias}, J. 1997, \aap, 322, 320

\end{thebibliography}

\begin{appendix}

\section{Detailed modelling procedure}
\label{app:detailed_modelling_procedure}
The modelling procedure is divided in two main steps. The first step uses non-rotating models to fit the seismic information of the target and constrain its location in the HR diagram. The second step uses rotating models and follows the HR track until the location identified by the first step to derive the corresponding rotation profile. This modelling procedure outputs a robust stellar structure and its corresponding rotation profile that can then be used to compute the rotational splittings and the asteroseismic rotation rate. Our targets are slow rotators, what motivated the use of this modelling strategy. Indeed, for slow rotators, it is possible to use a perturbative treatment of rotation, like for the Sun. In that case, the fit of the internal rotation can be separated from the fit of the structure, and the inferences about rotation are quasi-model-independent from the seismic structure at first order. In such framework, our modelling is appropriate and does not change the conclusions of our study. From a pure stellar modelling point-of-view, neglecting rotation in the first step is a simplification that has a small impact on the transport of the chemical elements. It can be associated with a systematic uncertainty that accounts for the uncertainties due to the choice of the physical ingredients (e.g. choice of abundances, opacities, diffusive formalism, etc.) (see JB22). The estimation of this systematic uncertainty is computationally expensive and was performed for some targets in the literature \citep[e.g.][]{Buldgen2019f,Farnir2020,Betrisey2022}. For these targets, a rich variety of physical ingredients changes were considered, and the resulting systematic uncertainty was smaller than the other sources of uncertainties.

The modelling procedure of the first step is similar to that of JB22. It consists in fitting the seismic information with a Markov Chain Monte Carlo (MCMC) in a grid of non-rotating models, coupled with a mean density inversion. We used a grid of non-rotating models computed with the Code Li\'{e}geois d’\'{E}volution Stellaire \citep[CLES,][]{Scuflaire2008b}, and the frequencies were computed with the Li\`{e}ge Oscillation Code \citep[LOSC,][]{Scuflaire2008a}. This grid is a high-resolution grid, consiting of 1.3 million models and whose specificities are summarised in Table \ref{tab:Spelaion_properties}. The physical ingredients are the same as in Sec. 2.1 of JB22. The minimisations were conducted with AIMS \citep{Rendle2019}, first by fitting the individual frequencies and the classical constraints (effective temperature, metallicity, and luminosity or frequency of maximal power $\nu_{max}$). Then, a mean density inversion \citep{Reese2012} is performed to constrain the mean density \citep[see e.g.][for a review about inversion techniques]{Buldgen2022_Fr} that is added to the set of classical constraints, assuming a conservative precision of 0.6\%, for a second MCMC fitting this time frequency separation ratios \citep[$r_{01}$ and $r_{02}$,][]{Roxburgh&Vorontsov2003} instead of the individual frequencies. We used uniform priors on the estimated stellar parameters. The likelihoods were computed assuming the true value of the observations were perturbed by some normally-distributed random noise (see JB22 for further details).

\begin{table}[h!]
\centering
\caption{Mesh properties of the grid used for the minimisation in the first step of the modelling strategy.}
\begin{tabular}{lccc}
\hline 
 &  Minimum & Maximum & Step \\ 
\hline \hline 
Mass $(M_\odot)$ & 0.80 & 1.60 & 0.02 \\  
$X_0$ & 0.67 & 0.74 & 0.01 \\  
$Z_0$ & 0.008 & 0.030 & 0.001 \\ 
\hline  
\end{tabular}
\label{tab:Spelaion_properties}
\end{table}

Rotating models are then computed based on the stellar properties determined from this asteroseismic modelling. In addition to the internal transport of AM described in the main text, these rotating models account for braking of the surface due to magnetised winds following \cite{mat15} with the torque:
\begin{eqnarray}
\nonumber
 \frac{{\rm d} J}{{\rm d}t} = \left\{
\begin{array}{l l }
-T_{\odot} \left({\displaystyle \frac{R}{R_\odot}} \right)^{3.1} 
\left({\displaystyle \frac{M}{M_\odot} }\right)^{0.5} \left({\displaystyle \frac{\tau_{\rm cz}}{\tau_{{\rm cz}\odot}} }\right)^{p} \left({\displaystyle \frac{\Omega}{\Omega_{\odot}} }\right)^{p+1} & 
(Ro > Ro_{\odot}/\chi) \\
\nonumber
 & \\
\nonumber
-T_{\odot} \left({\displaystyle \frac{R}{R_\odot}} \right)^{3.1} 
\left({\displaystyle \frac{M}{M_\odot} }\right)^{0.5} \chi^{p} \left({\displaystyle \frac{\Omega}{\Omega_{\odot}} }\right)
& (Ro \leq Ro_{\odot}/\chi) \; .  
 \end{array}   \right.
\end{eqnarray}

$R$ and $M$ are the radius and mass of the star, $Ro$ is the Rossby number and $\tau_{\rm cz}$ is the convective turnover timescale. Following \citet{Eggenberger2019}, the transition from the saturated to the unsaturated regime is defined with a parameter $\chi=Ro_{\odot}/Ro_{\rm sat}$ fixed to 10, the coefficient $p$ is fixed to 2.3 and a solar-calibrated braking constant $T_{\odot}$ is used.

\section{Supplementary data}
\label{app:supplementary_data}

\begin{figure}[h!]
\centering
\begin{subfigure}{.45\textwidth}
  \includegraphics[width=.99\linewidth]{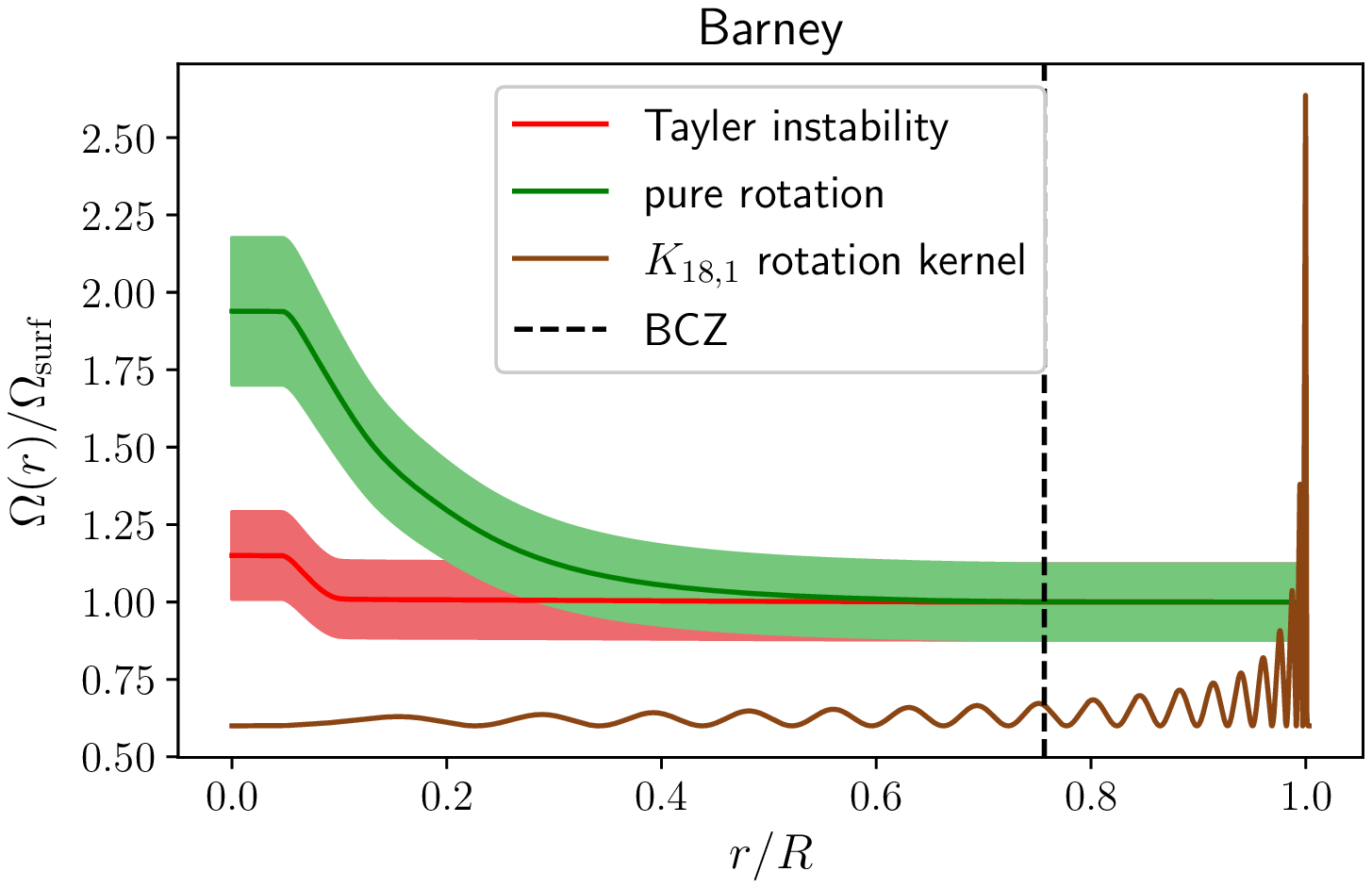}  
  \label{fig:barney_rotation_profiles}
\end{subfigure}
\begin{subfigure}{.45\textwidth}
  \includegraphics[width=.99\linewidth]{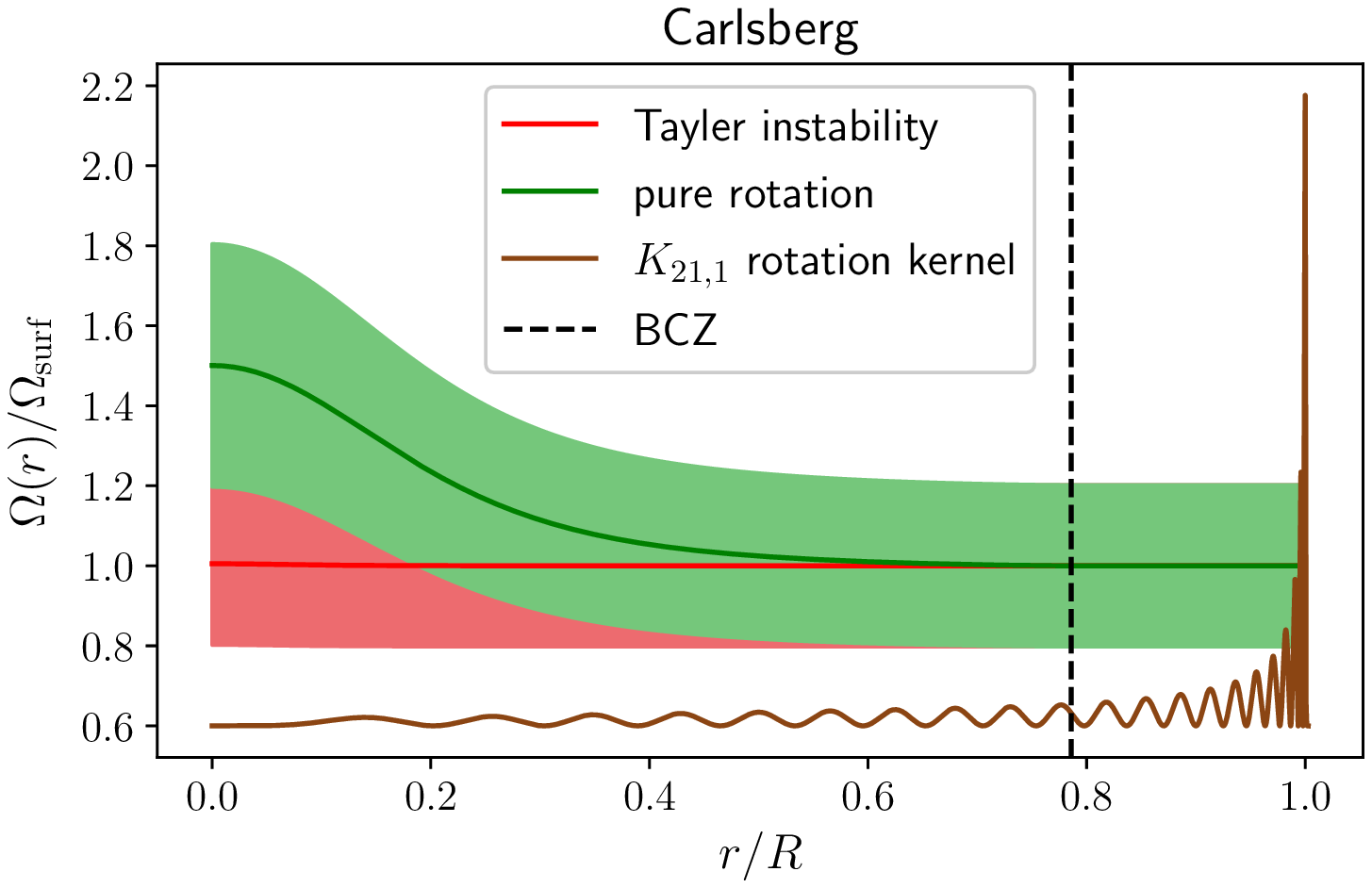}  
  \label{fig:Carlsberg_rotation_profiles}
\end{subfigure}
\caption{\textit{Left:} Rotation profiles of Barney (upper panel) and Carlsberg (lower panel). The base of the convective zone (BCZ) is show in dashed black. The rotation kernel is shown in brown, with $l=1$ and $n=18$ or $n=21$ to correspond to a frequency around the $\nu_{max}$ respectively for Barney and Carlsberg, and is rescaled and shifted vertically for illustration purposes.}
\label{fig:appendix:rotation_profiles}
\end{figure}

\begin{figure}[t!]
\centering
\begin{subfigure}{.45\textwidth}
  \includegraphics[width=.99\linewidth]{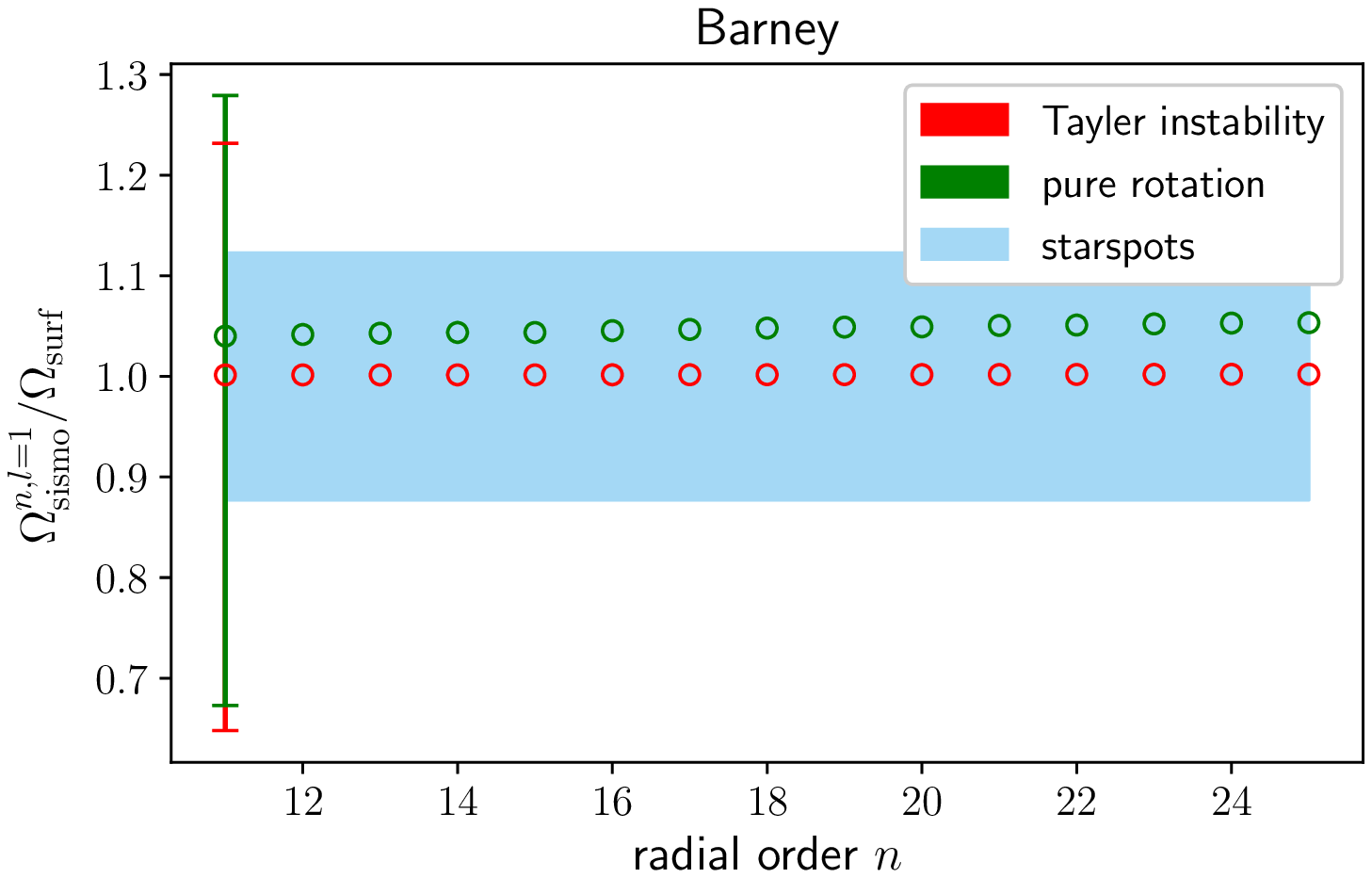}  
  \label{fig:Barney_sn11}
\end{subfigure}
\begin{subfigure}{.45\textwidth}
  \includegraphics[width=.99\linewidth]{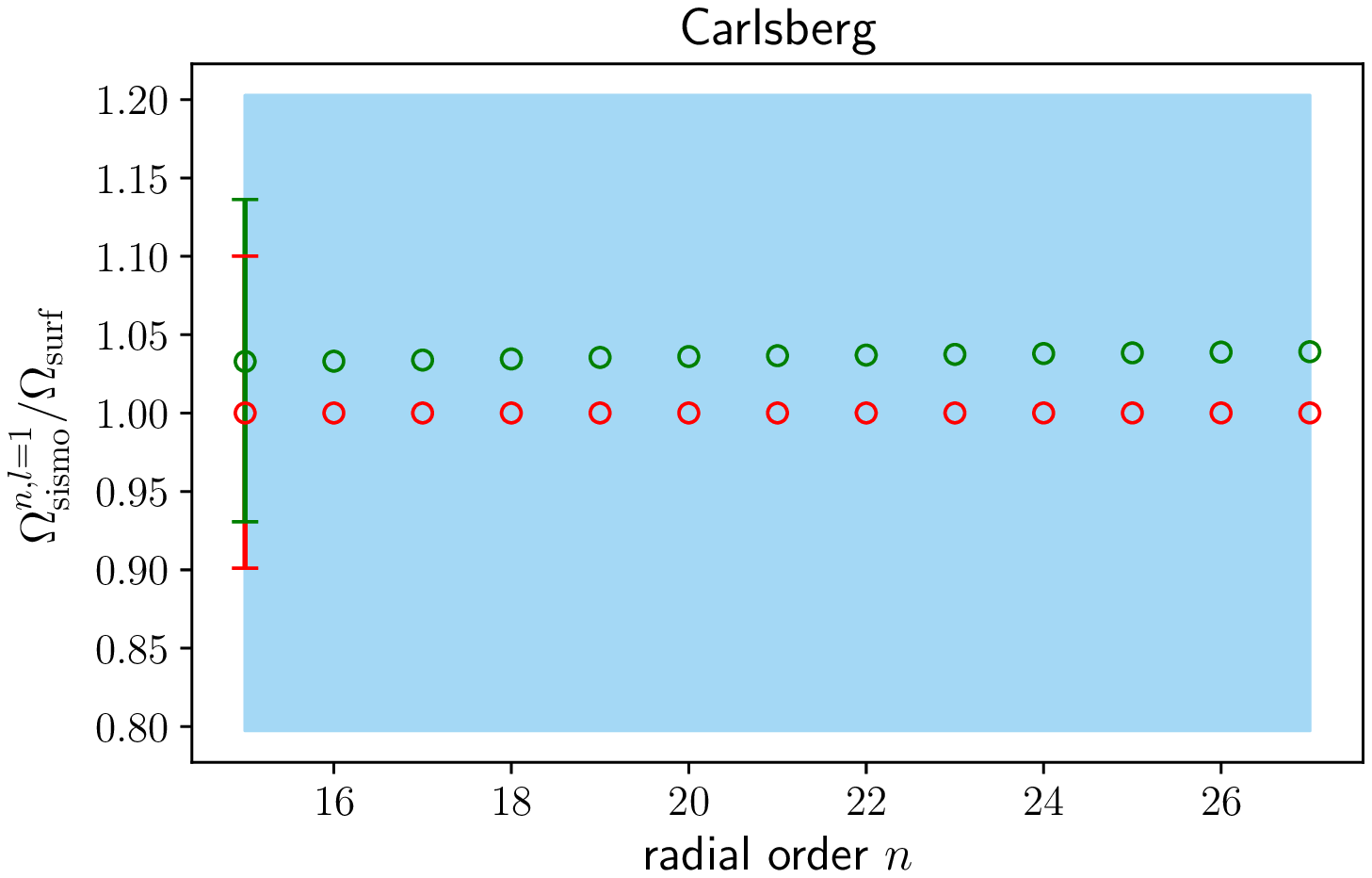}  
  \label{fig:Carlsberg_sn11}
\end{subfigure}
\caption{Surface rotation predicted by the seismology for Barney (upper panel) and Carlsberg (lower panel). The error bar corresponds to the precision of the average observational rotational splitting measured by OB18.}
\label{fig:appendix:splittings}
\end{figure}

In Fig. \ref{fig:appendix:rotation_profiles}, we show the rotation profiles of the different models of Barney and Carlsberg, and in Fig. \ref{fig:appendix:splittings}, we show the corresponding ratio between the asteroseismic and surface rotation rates.

\end{appendix}
\end{document}